\documentclass[letter]{jpsj3}
\usepackage{bm}
\usepackage{color}

\def\beq{\begin{equation}}
\def\eeq{\end{equation}}
\def\beqa{\begin{eqnarray}}
\def\eeqa{\end{eqnarray}}
\def\adag{a^{\dagger}}
\def\bdag{b^{\dagger}}

\def\degC{\kern-.2em\r{}\kern-.3em C}

\begin{document}

\title{
Instantaneous Intervalley Transition just at the Franck-Condon State \\ in the Conduction Band of GaAs 
}

\author{Hiromasa Ohnishi$^1$\thanks{hohnishi@tsuruoka-nct.ac.jp}, Norikazu Tomita$^2$, and Keiichiro Nasu$^3$}
\inst{$^1$National Institute of Technology (NIT), Tsuruoka College, 104 Sawada, Inooka, Tsuruoka, 
Yamagata, 997-8511,Japan\\
$^2$Department of Physics, Yamagata University, 1-4-12 Kojirakawa, Yamagata, 
990-8560, Japan\\
$^3$Institute of Materials Structure Science, High Energy Accelerator Research
Organization (KEK), 1-1 Oho, Tsukuba, 305-0801, Japan} 

\abst{
We propose two possible quantum microscopic mechanisms 
for intervalley transitions just at the Franck-Condon state in GaAs.
The first is the simultaneously combined coulombic elastic transitions of two electrons,
from the originally photo-excited $\Gamma$ valley to the mutually opposite L and -L valleys
in the Brillouin zone. 
The second is the ``elastic" scattering of an electron from the $\Gamma$ valley to the L one
by the frozen phonon, 
which is not newly created in the excited state, but already exists, being inherited from the starting ground state,
according to the Franck-Condon principle of the photo-excitation.
Surprisingly,
we can theoretically show that such the elastic electron-phonon scattering  gives extremely 
fast time constant of 
the order of a few tens femtoseconds, as well as the above elastic electron-electron scattering does.
}


\maketitle


Dynamics of photo-excited electrons and phonons in various crystals
is one of the most topical issues in the materials science
from both academic and engineering points of view. 
Nowadays, we can obtain time-resolved spectral data, 
within such a short laser duration period as a few tens femtosecond (fs). 
In some cases, an even more shorter time resolution becomes possible \cite{atto}. 
Recent developments of the experimental techniques also make us possible 
to obtain both momentum and energy resolved spectral data  
simultaneously and quite finely \cite{tanimura}. 
Thus, we can directly access to``time-evolving" spectral dynamics of electrons and phonons 
in crystalline materials in detail, under the explicit viewpoint of 
energy and momentum conservation laws.
These developments of the experimental techniques 
bring us remarkable and novel concepts on variety of phenomena, 
such as the photo-induced phase change\cite{nasu,pimit,ohkoshi} and 
the atomic scale tunnelling \cite{tunnel},
which are closely related to future device application.

Carrier relaxation dynamics in semiconductors is one of such phenomena:
So far, the intervalley scattering (IVS) in the conduction band of GaAs has been studied intensively,
and the fundamental understanding is established
with good agreement between theory and experiments, as described later.
On the other hand, the time-resolved optical measurement gives us indirect information for the phenomenon
in many cases, and hence there is an ambiguity on the interpretation  for the measurement.
The recent time- and momentum-resolved 
two-photon photo-emission spectroscopy (2PPES),  however, 
can evaluate the energy distribution function in the conduction band directly
without the aforementioned ambiguity.
This is performed by Kanasaki \textit{et. al.} for the first time, and 
they have succeeded to measure the $\Gamma$ to L ($\Gamma$-L) IVS, and 
determined that this intervalley transition occurs within such a short time 
as 20 fs, by directly observing substantial numbers of electron population 
at the L valley \cite{tanimura}.  

The dynamics of photo-excited electrons and phonons in early stage are understood 
in the context of the Franck-Condon principle, 
in which the excitation due to visible light absorption completes so suddenly that,
immediately after this excitation, only the electrons 
in the whole electron-phonon (e-ph) coupled system 
can change or move,
while the phonons are still in the starting ground state configuration.
At this moment, the electronic system is in a highly non-equilibrium state,
and the energy is gradually released into the phonon system through the e-ph interaction.
Afterwards, the phonon system is excited and starts to oscillate.
So far, the IVS in GaAs is considered on this scheme with the inelastic e-ph scattering process
\cite{tang,shank, shank2,rossi,gaasreview,mctrans,damen,sarma,cardona1,cardona2,stanton,saeta,qk,sjakste}.
In this mechanism,  the scattering e-ph complex created just after the excitation
must be separated into independent electrons and phonons to achieve a convergence of the electron into the final L valley.
For this purpose,
the phonon should recoil and dissipate away
from plane-wave electrons \cite{slow}.  
Then, the time constant for this $\Gamma$-L IVS is characterized 
by the phonon oscillation period, which is $\sim$ 110 fs \cite{qk}.
This scheme has been successfully applied for the phenomena so far,
but the aforementioned 2PPES measurement apparently gives one order shorter time constant 
than the previous results.
This aspect implies a possibility that there exists a novel mechanism 
for the IVS  ``just at the Franck-Condon state" in GaAs. 

\begin{figure}[tbp]
\includegraphics[width=8cm]{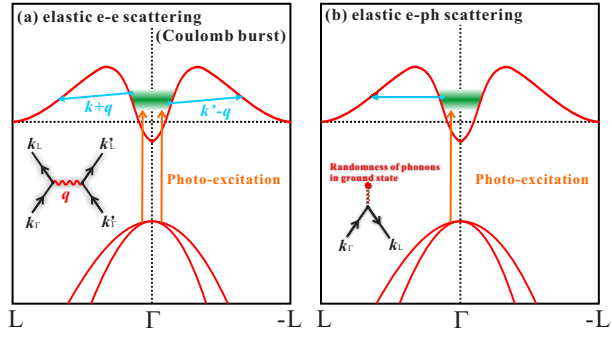} 
\caption{(Color online) Schematic explanations of (a) elastic e-e scattering (Coulomb burst) and (b)
elastic e-ph scattering are shown.}
\label{fig1}
\end{figure}

In this Letter,
we discuss a extremely rapid IVS mechanism experimentally observed in GaAs.
Especially, we focus on the very early stage of the phenomenon,
in which the phonon is frozen at the starting ground state configuration.
We theoretically propose two possible instantaneous $\Gamma$-L IVS mechanisms 
just at this Franck-Condon state.  
The first is the simultaneously combined elastic coulombic transitions of two electrons,
from the originally photo-excited $\Gamma$ valley to the mutually opposite L and -L valleys in the Brillouin zone,
as schematically shown in Fig.\ref{fig1}a.
Since this type coulombic transition can occur along the independent four directions 
in the three-dimensional Brillouin zone of GaAs, 
it is just the multi-electron Coulomb burst from the $\Gamma$ valley at the zone center to the all L ones in the zone edge.
Although importance of electron-electron (e-e) coulombic interaction in the early stage of the phenomenon has already been suggested in Ref. \cite{eemc},
here we examine it again with a clear quantum mechanical and microscopic picture.

The second is the $\Gamma$-L ``elastic" scattering of an electron by the frozen phonon,
which is not newly created in the excited state but already exists, being inherited from the starting ground state,
according to the Franck-Condon principle (FCP) of the photo-excitation.
The phonon system gives only a frozen spatial randomness for electronic system,
which mixes the eigenstates of $\Gamma$ valley with those of L one.
Thus, the photo-injected electrons to the $\Gamma$ valley automatically diffuse into the L valley
even within a short time just after the excitation, regardless of the phonon frequency. 
This mechanism is basically the same one proposed in Ref. \cite{franck}.
As a result, the elastic $\Gamma$-L transition is realized, as schematically shown in Fig.\ref{fig1}b.

We will theoretically show that the above two mechanisms can give extremely fast time constant 
of the order of a few tens fs.
They are not exclusive with each other, and can coexist in GaAs.
Furthermore, they are clearly distinguished since the thermal randomness or the mean amplitude of 
the frozen phonon in the starting ground state has an ordinary temperature dependence, 
while the e-e scattering has not.

As a standard model to represent the system having e-e and e-ph interactions, 
we employ the following Hamiltonian($\equiv H$), 
\beq
H= H_0 +H_{\rm I}, H_0= H_{\rm e}+H_{\rm p}, H_{\rm I} = H_{\rm ee}+H_{\rm ep},
\label{hamall}
\eeq
\beq
H_{\rm e}=\sum_{{\bm k},\sigma} E({\bm k}) a^\dagger_{{\bm k},\sigma} a_{{\bm k},\sigma},
H_{\rm p}=\sum_{\bm q} \omega_{\bm q} b^\dagger_{{\bm q}} b_{{\bm q}},
\label{ham0}
\eeq
\beq
H_{\rm ee}=U \sum_{\bm l} n_{{\bm l},\uparrow} n_{{\bm l}, \downarrow},
\label{hamee}
\eeq
\beq
H_{\rm ep}=(2N)^{\frac{1}{2}} \sum_{{\bm q},{\bm k},\sigma} S_{\bm q}
(b^{\dagger}_{\bm q}+b_{-{\bm q}}) a^{\dagger}_{{\bm k}-{\bm q},\sigma} a_{{\bm k},\sigma}.
\label{hamep}
\eeq
Here, $E({\bm k})$ is the one electron energy
in the conduction band.
$\adag_{{\bm k},\sigma}(a_{{\bm k},\sigma})$ is creation (annihilation) operator of 
an electron with wavevector ${\bm k}$ and spin $\sigma$.
$\bdag_{\bm q}(b_{\bm q})$
is creation (annihilation) operator of  a phonon,
with a wavevector ${\bm q}$.
$\omega_{\bm q}$ and $S_{\bm q}$ are the phonon energy and the e-ph coupling
constant, respectively.
$N$ is the total number of lattice site.
 $n_{{\bm l},\sigma}\equiv \adag_{{\bm l},\sigma}a_{{\bm l},\sigma}$ where
$a_{ {\bm l},\sigma}\equiv N^{-1/2}\sum_k e^{i{\bm k} \cdot {\bm l}} a_{{\bm k}, \sigma}$.
The unit of length is the lattice constant along [111] direction of the GaAs crystal.
The wave number dependence of coulombic interaction among electrons are neglected,
by considering only the onsite coulombic repulsion $U$.
This is because we consider only $\Gamma$-L scattering process, 
which is dominated by ${\bm q} \sim \pi$ component. 

The density matrix $(\equiv\rho(t))$ at a time $t$ is written as a direct product of
the electron density matrix $(\equiv\rho_{{\rm\scriptsize e}} (t))$ and the phonon one 
$\rho_{{\rm\scriptsize p}} \equiv \exp{(-H_{{\rm\scriptsize p}}/k_{{\rm\scriptsize B}} T)}$
with phonon temperature $T$ .
Our electronic system is in a non-equilibrium state starting from the photo-excitation.
Time evolution of the density matrix is obtained by 
\begin{equation}
\tilde{\rho}(t+\Delta t)=\exp_+ \left\{ i\int^{\Delta t}_0 d\tau \tilde{H}_{{\rm\scriptsize I}}(\tau) \right\}
\rho_e(t)\rho_p
\exp_- \left\{ -i\int^{\Delta t}_0 d\tau' \tilde{H}_{{\rm\scriptsize I}}(\tau') \right\},
\label{rhodef}
\end{equation}
where subscript $\pm$ in exponent means chronological order and $\tilde{O}(\tau) \equiv e^{i\tau H_{{\rm\scriptsize 0}}} O e^{-i\tau H_{{\rm\scriptsize 0}}} $.
Time evolution of $n_{\bm{k},\sigma}(t)$ is evaluated from
\beq
n_{{\bm k}, \sigma} (t) \equiv \langle n_{{\bm k}, \sigma} (t) \rangle 
= \frac{ {\rm Tr} (n_{{\bm k}, \sigma} \tilde{\rho}(t)) }{{\rm Tr}(\tilde{\rho}(t))},
\label{expand}
\eeq
by expanding the exponential terms in Eq. (\ref{rhodef}) 
up to the second order of $H_{{\rm\scriptsize I}}$.

Since the following calculation is straightforward, 
we show the procedure only for a gain channel by $H_{\rm ep}$, which is
\beq
\frac{ {\rm Tr}\left( 
n_{{\bm k}, \sigma} \int^{\Delta t}_0 d\tau \tilde{H}_{{\rm ep}} (\tau)
\rho_{{\rm\scriptsize e}} (0) \rho_{{\rm ep}} 
\int^{\Delta t}_0 d\tau' \tilde{H}_{{\rm\scriptsize ep}} (\tau')
\right)
}{ {\rm Tr}( \rho_{{\rm\scriptsize e}} (0) \rho_{{\rm\scriptsize p}} ) }.
\label{hepdecay1}
\eeq
By changing the integration variables as $(\tau+\tau')/2$ and $\gamma\equiv \tau'-\tau$,
and considering the limit $\Delta t\to \infty$ (Fermi's golden rule),
we obtain
\beq
\frac{\Delta t \int^{\infty}_{-\infty} d\gamma
{\rm Tr} \left( 
\tilde{H}_{{\rm ep}} (\gamma)
n_{{\bm k}, \sigma} 
\tilde{H}_{{\rm ep}} (0)
\rho_{{\rm\scriptsize e}} (0) \rho_{{\rm ep}}
\right)
}{ {\rm Tr}( \rho_{{\rm\scriptsize e}} (0) \rho_{{\rm\scriptsize p}} ) } .
\label{expand2}
\eeq
It should be noted that, 
the time correlation function of the form of 
$\langle H_{\rm I}(\gamma)H_{\rm I}(0) \rangle$ would decay in the shorter time than that of the 
integration range in the present system, 
since the number of the photo-injected electrons is quite low 
compared with the number of states of the wide conduction band ($\sim$0.003 electrons/sites) \cite{tanimura},
implying the conduction band itself works as a heat reservoir for the photo-injected electrons.
Then, 
the use of the Fermi's golden rule is still validated for the present process.

By inserting the Eq. (\ref{hamep}) to above equation, 
we obtain the transition rate.
At this stage, we explicitly take into account the FCP of the photo-excitation,
in which only the electrons in the whole e-ph coupled system can change or move,
while the phonons are still in the starting ground state configuration.
This situation is reflected by considering only the zeroth order term of the phonon time evolution as
\begin{align}
\tilde{b}^{(\dagger)}_{{\bm q}}(t)=& e^{(-)i \omega_{\bm q} t} b^{(\dagger)}_{{\bm q}}(0) \notag \\ 
=&b^{(\dagger)}_{{\bm q}}(0) \left\{
1+(-)i\omega_{\bm q}t-\omega^2_{\bm q}t^2+\cdots
\right\} \notag \\
=& b^{(\dagger)}_{{\bm q}}(0)\left\{ 
1+{\rm O} \left(\frac{\omega_{\bm q}}{\rm Band~Width}\right)\right\},
\label{phevolv}
\end{align}
while $\tilde{a}^{(\dagger)}_{{\bm k}, \sigma}(t)=e^{(-)i E({\bm k})t} a^{(\dagger)}_{{\bm k}, \sigma}(0)$ for electrons.
After a tractable calculation, finally we get the transition rate,
including electron and phonon numbers as,
\begin{equation}
\Gamma^{\rm g}_{{\rm ep},{\bm k},\sigma} (T) \equiv 
\frac{2\pi }{N} \sum_{\bm q} S^2_{\bm q}
(\langle \langle n_{\bm q} (T) \rangle \rangle +\frac{1}{2}) 
(1- \langle n_{{\bm k},\sigma}  \rangle)
\langle n_{{\bm k}+{\bm q}, \sigma} \rangle \delta ( E({\bm k})-E({\bm k}+{\bm q})),
\label{gamepg}
\end{equation}
where $\langle \langle n_{\bm q} (T) \rangle \rangle=(\exp{(\omega_{\bm q}/k_{\rm B}T)}-1)^{-1}$
is the phonon number at the FC state.
If the time evolution of phonon is fully taken into account at Eq. (\ref{phevolv}),
the transition rate by the ordinary inelastic e-ph scattering is obtained \cite{mctrans}.  

By considering all the channels, we obtain the rate equation for electronic population as
\begin{equation}
\frac{\partial n_{{\bm k}, \sigma} (t)}{\partial t}=
\left\{ \Gamma^{\rm g}_{{\rm ee}, {\bm k}, \sigma}+\Gamma^{\rm g}_{{\rm ep}, {\bm k}, \sigma}(T) \right\}
- \left\{ \Gamma^{\rm d}_{{\rm ee}, {\bm k}, \sigma}+\Gamma^{\rm d}_{{\rm ep}, {\bm k}, \sigma}(T) \right\},
\label{rateeq}
\end{equation}
where the superscripts g and d represent gain and decay channels, respectively.
Since we focus on the $\Gamma$-L process very soon after the photo-excitation,
we consider only the decay channel of the $\Gamma$-valley electrons, which is photo-injected.
The corresponding transition rate is given as
\begin{equation}
\Gamma^{\rm d}_{{\rm ep}, {\bm k}, \sigma}(T)
=\frac{2\pi }{N} \sum_{\bm q} S^2_{\bm q}
(\langle \langle n_{\bm q} (T) \rangle \rangle +\frac{1}{2})
(1- \langle n_{{\bm k}+{\bm q}, \sigma} \rangle)
\langle n_{{\bm k},\sigma} \rangle \delta ( E({\bm k}+{\bm q})-E({\bm k})).
\label{gamep}
\end {equation}

\begin{figure}[tbp]
\includegraphics[width=7cm]{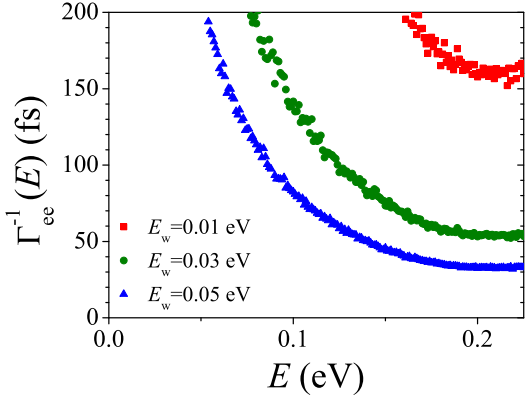} 
\caption{\label{fig2}(Color online) Time constant by elastic e-e scattering ($\Gamma^{-1}_{\rm{ee}}$) is given.
The energy is referenced from that of L valley bottom.}
\end{figure}
\begin{figure}[tbp]
\includegraphics[width=7cm]{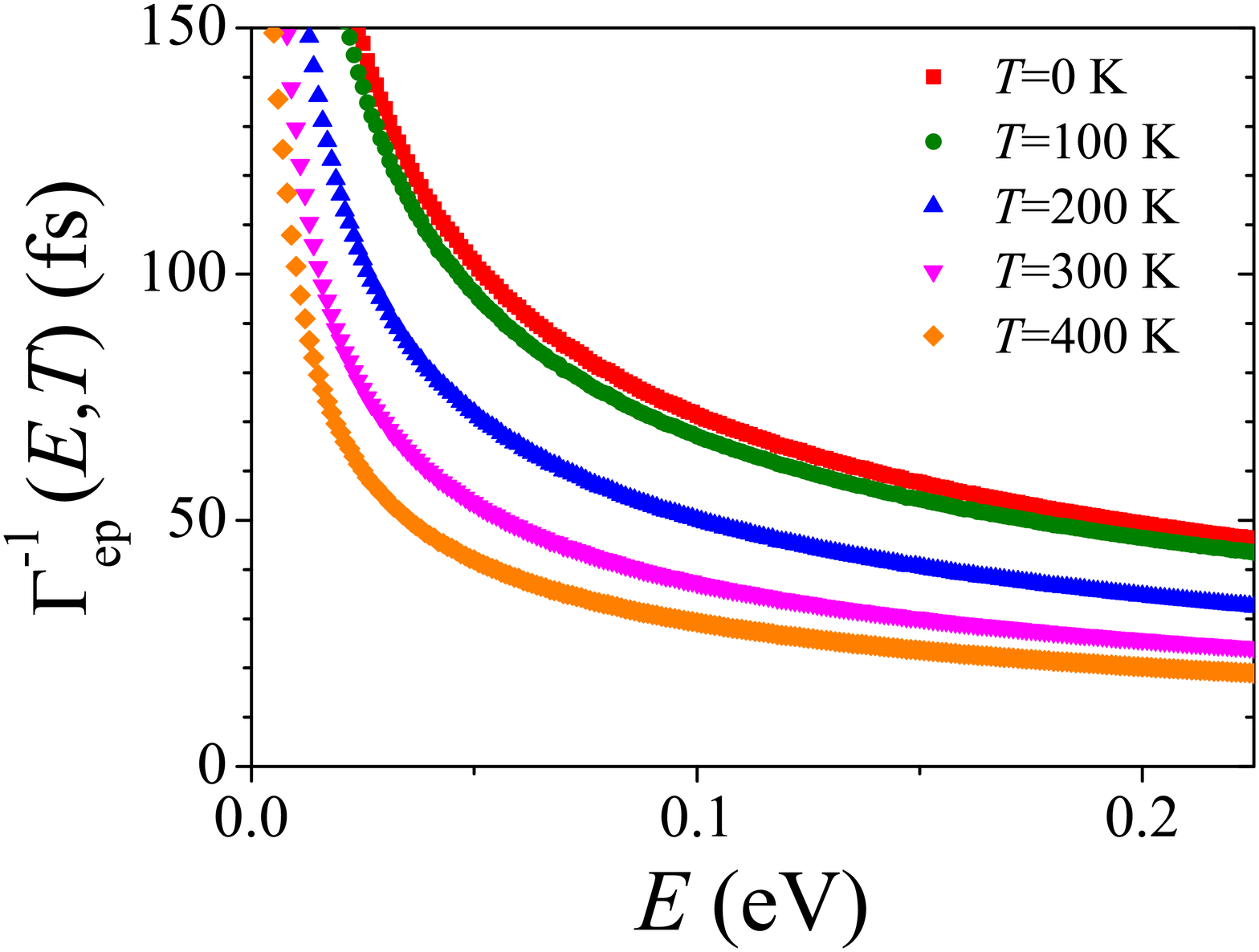} 
\caption{\label{fig3}(Color online) Time constant 
by elastic e-ph scattering  ($\Gamma^{-1}_{\rm{ep}}$) is given.
The energy is referenced from that of L valley bottom.}
\end{figure}

We also assume that the initial $\Gamma$-valley is fully occupied and 
the L-valley is completely unoccupied. 
Then, the transition rate by the e-ph scattering for the $\Gamma$-L process is given as
\begin{equation}
\Gamma^{\rm d}_{{\rm ep}, {\bm k}, \sigma}(T)=\frac{2\pi }{N} \sum_{\bm q} S^2_{\bm q}
(\langle \langle n_{\bm q} (T) \rangle \rangle +\frac{1}{2}) 
\delta ( E_L ({\bm k}+{\bm q})-E_{\Gamma} ({\bm k})).
\label{gamepf}
\end{equation}
where $E_{\Gamma(L)}$ represents the band energy in $\Gamma$(L)-valley.

Following the same manner, we also obtain the transition rate by the e-e scattering as
\begin{equation}
\Gamma^{\rm d}_{{\rm ee}, {\bm k},\sigma}= \frac{2\pi U^2}{N^2} \sum_{{\bm k'},{\bm{q}}} 
\delta (E_L ({\bm k}+{\bm q})+E_L({\bm k'}-{\bm q})-E_{\Gamma}({\bm k})-E_{\Gamma}({\bm k'})).
\label{gameef}
\end{equation}

By noting four sets of L and -L valleys exist in the Brillouin zone,
the energy dependent transition rate is given as
\beq
\Gamma_{{\rm ee(ep)}} (E)=4 \sum_{{\bm k} \subseteq {\bm k}(E)}
\Gamma^{\rm d}_{{\rm ee(ep)}, {\bm k},\sigma}/N_s,
\label{gamsum}
\eeq
where $N_s$ is the corresponding number of states. 
The time constant is obtained as inverse of the transition rate.
It should be noted that, in our model, 
the higher order multi-electrons scattering process, 
which is able to reduce to two-electrons scattering,
is included as a linear summation of the second order perturbation process.
The same argument also holds for the e-ph scattering.

To numerically evaluate above time constants,
Eqs. (\ref{gamepf})-(\ref{gamsum}) are calculated, 
by replacing the discretized $\bm{k(q)}$ 
to continuous one, as 
$N^{-1}\sum_{\bm{k(q)}} \rightarrow (4\pi^4/3)^{-1}\int d{\bm{k(q)}}$.
For the sake of simplicity, 
one electron energies, $E_{\Gamma}({\bm k})$ and $E_{L}({\bm k})$, 
are effectively parametrized only around the L valley bottom by following functions, 
from Ref. \cite{band0} as
\beq
E_{\Gamma}({\bm k})=c_1 \left[ \left\{ \frac{{\bm k}\cdot{\bm k} }{c^2_2\pi^2}\right\}^{c_3} -1 \right],
\label{egam}
\eeq
\beq
E_{L}({\bm k})=c_4 \left( 1+\cos{k_L} \right)+\frac{1}{4} \frac{c_1}{c^2_2 \pi^2} 
\left( k^2_S + k^2_K \right),
\label{el}
\eeq
where $k_L$ is the wave vector directed parallel to $\Gamma$-L line, 
$k_S$ and $k_K$ are the wave vectors directed normal to $k_L$, respectively.
We set that $c_1=0.386$, $c_2=0.0862$, $c_3=0.3635$, and $c_4 =0.429$, where
the energy is referenced from that of the L valley bottom.
Although curvature of the band dispersion slightly depends on 
the method even in the \textit{ab initio} calculation \cite{band1,band2},
the basic feature is the same and the present result would not be altered significantly.
The delta function in the integral is replaced by the Gaussian one with the width of $0.01$ eV.
We set $U=5$ eV. This value is empirical one, 
which is estimated from the experimental exciton binding energy \cite{excitonbe}, 
and hence various higher order correlations and screening effects are already included, 
and that verifies the second order perturbation theory.
The e-ph scattering is dominated by the ${\bm q}\sim \pi$ components,
and thus we neglect the ${\bm q}$-dependence of $S_{\bm q}$ and 
$\omega_{\bf q}$, and set 
$S_{\bm \pi}=0.15$ eV, and $\omega_{\bf \pi}=30$ meV. 
This e-ph coupling constant is almost same as that of the standard deformation potential theory
\cite{deformation1,deformation2},
being not so strong as compared with the band width. Thus, the second order perturbation theory
will be enough. 
Multi-dimensional integration is performed 
by Monte Carlo (pseudo-random) integration technique. 
The samplings are performed more than $10^6$ times for Eqs. (\ref{gamepf}) and (\ref{gameef}),
and more than $10^5$ times for Eq. (\ref{gamsum}).
These are enough to get well converged values.

The result for the elastic e-e scattering is given in Fig. \ref{fig2}.
In this calculation, 
we assume that two electrons Coulombic scattering is possible only for 
the electrons in the $\Gamma$-valley
satisfy $\mid E_{\Gamma}({\bm k})-E_{\Gamma}({\bm k'}) \mid \le E_{\rm w}$,
since the photo-injected electrons are initially distributed 
energitically narrow region \cite{tanimura}.
Thus, $E_{\rm w}$ represents the initial distribution width of the photo-injected electrons.
The obtained time constant
depends on $E_{\rm w}$ because the number of simultaneously combined two electrons
are simply proportional to $E_{\rm w}$. 

In case of the elastic e-ph scattering, as seen in Fig. \ref{fig3},
the time constant strongly depends on the phonon temperature.
This is because thermal randomness of the frozen phonon in the starting ground state
has a ordinary temperature dependence, as aforementioned.
The elastic e-ph scattering surprisingly gives the faster time constant than the elastic e-e scattering.
Thus, the elastic e-ph scattering process would be a possible IVS mechanism in very early stage of the phenomenon.
On the other hand, it would be reasonable that the elastic e-e scattering also works at this stage.
Actually, experimentally observed temperature dependence \cite{tanimura} would be explained by 
the coexistence of both processes.  
 
It should be noted again that
the irreversibility of above elastic transitions comes from the wideness of the conduction band, 
which is about 5 eV or so. If we assume that our exciting photon has a small energy width,
final electron beams at around the $\Gamma$ point and the L one are the wave packets,  
propagating in the crystal with variously different three-dimensional vectorial 
group velocities. 
Since we have taken the lattice constant as the unit of length, 
magnitudes of these various vectorial velocities are of the order of band width, resulting in large irreversibility.
This situation is quite similar to the well-known Fano resonance \cite{fano}.

In summary, we have demonstrated elastic e-e and e-ph transition mechanisms for extremely rapid $\Gamma$-L IVS 
``just at the Franck-Condon state" in GaAs, which can give the time constants of the order of a few tens fs.
These two mechanisms are inferred to coexist in GaAs, and 
give consistent results for the experiment \cite{tanimura}.

%
%
\begin{acknowledgment}
The authors would like to thank K. Tanimura, J. Kanasaki, H. Tanimura for
presenting their research result prior to publication.
This work was partly supported by JSPS Grant-in-Aid 
for Specially Promoted Research, Grant Number 24000006.
\end{acknowledgment}


\end{document}